\documentstyle{article}

\input{tcilatex}

\begin{document}

\author{Anthony Rizzi \\
California Institute of Technology, LIGO, \\
Livingston, LA 70754}
\title{Angular and Linear Momentum in General Relativity: Their Geometric Structure
and Interrelation}
\date{}
\maketitle

\begin{abstract}
Generalized definitions for angular and linear momentum are given and shown
to reduce to the ADM (at spatial infinity) definitions and the definitions
at null infinity in the appropriate limit. These definitions are used to
express angular momentum in terms of linear momentum. The formalism allows
one to see the connection with the classical and special relativitistic
notions of momenta. Further, the techniques elucidate, for the first time,
the geometric nature of these conserved quantities. The boosted
Schwarzschild solution is used to illustrate some aspects. The definitions
are useful and give insight in the region far from all masses where gravity
waves are detected.

\end{abstract}

\section{Introduction}

\bigskip The key conserved quantities (energy, linear momentum and angular
momentum) are far from intuitive in general relativity. The non-locality of
the gravitational field, the lack (in general terms) of symmetries, and the
general coordinate invariance of the equations make such definitions
non-trivial.

Linear and angular momentum are the most subtle of the three conserved
quantities. Linear momentum was not defined at null infinity\footnote{%
``Null infinity'' is the place and time at which gravity waves arrive far
from all matter in an asymptotically flat space-time (cf. \cite
{Rizzi:NullgoestoSpatial}). The phrase ``null infinity'' arises out of the
``conformal'' picture where one ``brings'' infinity into a finite place by
the appropriate division in the limit.} until the 1960's (\cite
{Bondi-Burg-Metzner'62},\cite{Sachs'62},\cite{Sachs-VIII'62}) and angular
momentum waited until the late 1990's to get such a definition (cf. \cite
{RizziAMPRL},\cite{Rizzi:NullgoestoSpatial},\cite{RizziMG8},\cite
{RizziThesis}. Previous to these definitions only the spatial (ADM \cite
{ADM'60-61},\cite{ADM'62},\cite{York1980},\cite{Wald'84},\cite{York78})
definition of these quantities were known; that is, only definitions that
give the ``total momentum'' for the whole space-time were available. The
definitions at null infinity are particularly valuable at present because
they allow the exchange of momentum due to gravitational radiation, and
experimental science, e.g. in the form of LIGO or VIRGO, is rapidly moving
to empirically investigate gravitational radiation and its sources.

Here I give the generalized definitions of linear and angular momentum and
show that they reduce to the proper form for the ADM (spatial) and null
definitions. Furthermore, I give, for the first time also, the geometric
interpretation of both these quantities. In this context, it is shown that
light rays ``at'' null infinity can be used to probe the structure of the
space-time. Since the space-time structure results from a source at a given
retarded time one can thereby obtain the momenta of the source at that time.
In addition, this paper connects the definitions in GR with our classical
and special relativistic understanding. The calculations are in themselves
instructive because they reveal something of the facility and of the
mechanics of using the Christodoulou-Klainerman formalism which is so useful
for far field problems involving gravity waves.

In section two, we setup the foliation of space-time needed to facilitate
the analysis. In section three, we generalize the notion of curvature with
an eye toward generalizing the momenta; we take inspiration from the result
given in \cite{Rizzi:NullgoestoSpatial}. Section four, gives the decay law
for the extrinsic curvature. Section five, generalizes the linear momentum.
Section six shows that it is the correct generalization and reminds us that 
\cite{Rizzi:NullgoestoSpatial} shows that the angular momentum can be
written using the generalized curvature in the manner shown there. Finally,
the last section (seven) discusses the geometrical insights gained by the
process and the interrelation between the linear and angular momentum. This
section uses the Schwarzschild space-time to illustrate the point.

\section{The Foliation}

We begin by deciding on the foliation of a generic asymptotically-flat \cite
{C&k'93} space-time that will be needed to do the calculations in the
subsequent sections. The foliation will give us the surfaces of integration
we need for the definitions and facilitate taking the appropriate limits to
null and spatial infinity. We will use the foliation defined in \cite
{RizziAMPRL} (for more foliations confer \cite{C&k'93},\cite{Christo-long}, 
\cite{ChristoPRL},\cite{Rizzi:NullgoestoSpatial} and \cite{RizziThesis}).
The description of the foliation given in \cite{RizziAMPRL} is repeated here
for the convenience of the reader. The foliation is defined pictorially in
figure \ref{FAffine Foliation}; it chops the space-time near $C_{s}^{-}\,\ $%
into topologically $S^{2}$ surfaces. The foliation is created by starting
with a maximal (one with traceless extrinsic curvature) hypersurface at some
time $t$, $\Sigma _{t}$, and encircling, far from the matter, all matter in
the slice with a surface of constant potential (i.e. use radius coordinate, $%
r_{laplace}$, obtained from a level surface of the solution of Laplace's
equation for an electric charge in the spatial hypersurface $\Sigma _{t}$).
As one moves farther and farther away from all mass, the radius coordinate $%
r_{laplace}$ defines more and more closely a sphere until finally, ``at
infinity,'' it is precisely so. Label one of these far away surfaces on $%
\Sigma _{t},$ $S_{-k}$. Light rays moving inward from $S_{-k}$ with initial
tangent vector $\underline{l}^{\prime }=T^{\prime }-N^{\prime }$, where $%
T^{\prime }$ is the time direction normal to $\Sigma $ and $N^{\prime }$ is
the normal in $\Sigma $ perpendicular to $S_{-k}$, form a cone (topology $%
S^{2}\times R$) called $C_{s}^{-}.$\ The tangents to the null rays define $%
\underline{l}$ on the cone. Define an affine parameter $u$ such that: $%
\underline{l}u=1$ and $u=-k$ on $S_{-k}$ with $k$ large. All the $S^{2}$
surfaces that foliate the cone are labeled by the same affine parameter, $s$%
, such that $ls=1$ and $s=r\equiv \sqrt{\frac{\text{Surface\thinspace Area
of }S_{0,s}}{4\pi }}$ that is, using the areal radius of the $u=0$ surface, $%
S_{0,s}$. Next, define the outgoing null vector, $l$, at each point on the
cone by requiring $\underline{l}\cdot l=-2$. For each constant $u$ surface
on $C_{s}^{-}$, $S_{u,s}$, use $l$ to send light rays into the space-time to
generate a new cone, $C_{u}^{+}$. The topologically $S^{2}$ surfaces given
by $s$ and $u$ constant constitute leafs of a foliation of space-time near $%
C_{s}^{-}$. $C_{s}^{-}$ becomes null infinity as the radius coordinate
defined by Laplace's equation goes to infinity.\footnote{%
Taking the limit makes all the previous construction precise. One also lets $%
S_{-k}$ $\rightarrow S_{-\infty }^{*}$ so that $-\infty <u<\infty $ on null
infinity. Further, note the foliation construction picks out an origin on
the initial spatial slice.} Hence, the foliation is a ``near null infinity''
foliation. Finally, on each point of each leaf, we put a set of
appropriately normalized null tetrads; two spatial vectors, $e_{A}$ where $%
A\in \{1,2\}$ and we already have two null vectors:$\;$ $e_{3}=\underline{l}%
\ $, $e_{4}=l$ (tangent vector of the appropriate geodesic of $C_{u}^{+}$).
The connection coefficients, $\Gamma _{abc}=e_{a}\cdot \nabla _{e_{c}}e_{b}$
where$\ $ $\nabla $ is the covariant derivative with respect to the
space-time metric $g_{\mu \nu }$ and $a,b,c\in \{1,2,3,4\},$ are called the
Christoffel symbols when a coordinate basis is used. With a null tetrad
system, they are called the null Ricci rotation coeffiecients $\gamma _{abc}$%
. For convenience the Ricci Coefficients are defined in appendix I.

In short, the affine parameters, $s$ and $u$ defined above serve to label
the topologically $S^{2}$ surfaces $S_{u,s}$; it is on such surfaces that
the energy and momentum integrals are to be taken.

\section{Generalized Curvature}

\smallskip We can now generalize the extrinsic curvature of a maximal slice $%
\Sigma _{t}$ to an extrinsic curvature more appropriate to the $S^{2}$
slicing of the foliation. We then will evaluate the important curvature
components in terms of Ricci rotation coefficients (cf.appendix I of this
paper and \cite{Rizzi:NullgoestoSpatial} for summary, see also. \cite{C&k'93}%
,\cite{ChristoPRL},\cite{RizziAMPRL}) for the general case (arbitrary
``lapse function,'' $a$) and for two specific cases. We, first, recall the
definition of the extrinsic curvature of a maximal slice $\Sigma _{t}$ : 
\begin{eqnarray}
k_{ij} &=&-(D_{i}T^{\prime },e_{j})  \label{extrinsic curvature} \\
where &:&i,\,j\in \{1,2,3\},\text{the spatial indices}  \nonumber \\
&&\text{and }D_{\mu }\text{ is the covariant derivative in the }\mu \text{
direction}
\end{eqnarray}

Next, recall that the standard null pair is: (\cite{C&k'93},\cite
{Rizzi:NullgoestoSpatial}) $l^{\prime }=T^{\prime }+N^{\prime }$ and $%
\underline{l}^{\prime }=T^{\prime }-N^{\prime },$ where $T^{\prime }$ is the
unit vector orthogonal to the maximal slice and $N^{\prime }$ is the unit
vector perpendicular to the $S^{2}$ slice within the maximal slice $\Sigma
_{t}.$

Using quantities more natural to the $S^{2}$ slicing, one can rewrite
equation \ref{extrinsic curvature}. One can write the bi-normal to the $%
S^{2} $ surface as ${\bf B}^{\prime }=\frac{1}{2}(l^{\prime }+\underline{l}%
^{\prime })$ and the vector $N^{\prime }=\frac{1}{2}(l^{\prime }-\underline{l%
}^{\prime })$ normal to this and normal to the spatial vectors $e_{A}$ in
the tangent space of the $S^{2}$ surface. Which allows one to generalize
equation \ref{extrinsic curvature} as: 
\[
k_{ij}^{\prime }=-(D_{i}{\bf B}^{\prime },e_{j}) 
\]
where we introduced the prime on the $k_{ij}$ to indicate the use of the
standard pair, $l^{\prime },\underline{l}^{\prime }$. For future, reference, 
$k^{g}$ refers to the $k$ curvature associated with the geodesic pair 
\footnote{%
To be complete and unambiguous, this null pair should be called the {\em %
affine} geodesic pair, because it has the ``normalization'' that makes the
tangent vector to the geodesic null ray correspond to an affine
parameterization of the geodesic. However, because the affine parameter is
the most natural to the null geodesic, we will use the term ``geodesic null
pair.''} ($l=a^{-1}\,l^{\prime },\underline{l}=a\,\underline{l}^{\prime }$
(also could be written $l^{g},\underline{l}^{g}$)) and $k^{t}$ refers to
that associated with the $t$-null pair ($l^{t}=\phi \,l^{\prime },$ $%
\underline{l}^{t}=\phi ^{-1}\underline{l}^{\prime }$). For instance, ${\bf B}%
^{t}=\frac{1}{2}(l^{t}+\underline{l}^{t})$ and $k_{ij}^{t}=-(D_{i}{\bf B}%
^{t},e_{j}).$

A brief aside can give an intuitive understanding of the meaning of the
lapse transformation and hence further manifest the need to use the
appropriate lapse function. Such insight comes from studying the lowest
dimensional special relativistic case. In 2-D Minkowski space-time, pick an
inertial frame, and establish the following natural coordinates $\{t,x\}$
(or equivalently, $\{t,r\}$). The standard null pair is then: $l^{\prime
}=\{1,1\},\,\underline{l}^{\prime }=\{1,-1\}.$ If we now boost to a
different frame, labeled by superscript $`\rightarrow ^{\prime }$, moving at
speed $\beta $, we have, in the new frame, $l^{\prime \rightarrow
}=a\,l^{\prime },\,\underline{l}^{\prime \rightarrow }=a^{-1}\,\underline{l}%
^{\prime }$ where $a=\sqrt{\frac{1-\beta }{1+\beta }.}$ In this way, it is
clear that use of this null pair in the boosted frame is inappropriate,
because it has the wrong lapse function.\footnote{%
A lapse transformation is necessary to get to obtain the null pair
appropriate to the boosted frame.} Its lapse function is the one appropriate
for the initial frame, because its binormal is the time direction in that
frame. Said another way, the choice of binormal, $\overrightarrow{{\bf B}}$,
is a choice rest frame with its time direction pointed along $%
\overrightarrow{{\bf B}}$, and in this frame, the tangent vectors to the
light rays emitted inward and outward at rest in that frame will give the
appropriate null pair (respectively, $l$ and $\underline{l}$) for that frame.

Now, we can write the general $k_{ij}$ in terms of the Ricci rotation
coefficients (cf. appendix I and \cite{C&k'93}, also cf appendix C of \cite
{Rizzi:NullgoestoSpatial}). Here, we use a superscript $``any"$ on $%
k_{ij}\,, $ $k^{any},$ to indicate the general nature of the calculation;
i.e. that we have not specified a particular $l$-pair. 
\begin{eqnarray*}
k_{AB}^{any} &=&-\left( D_{A}\left( \frac{1}{2}(l+\underline{l})\right)
,e_{B}\right) \\
&=&-\frac{1}{2}(\chi _{AB}+\underline{\chi }_{AB})
\end{eqnarray*}

\begin{eqnarray*}
k_{AN}^{any} &=&-(D_{A}{\bf B},N) \\
&=&-\frac{1}{4}\left( D_{A}(l+\underline{l}),(l-\underline{l})\right) \\
&=&-\frac{1}{4}\left( -(D_{A}l,\underline{l})+(D_{A}\underline{l},l)\right)
\\
&=&V_{A}
\end{eqnarray*}

\begin{eqnarray*}
k_{NN}^{any} &=&-\frac{1}{8}\left( D_{(l-\underline{l})}(l+\underline{l}),(l-%
\underline{l})\right) \\
&=&-\frac{1}{8}\left( -(D_{(l-\underline{l})}l,\underline{l})+(D_{(l-%
\underline{l})}\underline{l},l)\right) \\
&=&-\frac{1}{8}\left( -(D_{l}l,\underline{l})+(D_{\underline{l}}l,\underline{%
l})+(D_{l}\underline{l},l)-(D_{\underline{l}}\underline{l},l)\right) \\
&=&-\frac{1}{8}\left( -4\Omega -4\underline{\Omega }-4\Omega -4\underline{%
\Omega }\right) \\
&=&\Omega +\underline{\Omega }
\end{eqnarray*}

\begin{eqnarray*}
k_{{\bf B}N}^{any} &=&-(D_{{\bf B}}{\bf B},N) \\
&=&-\frac{1}{8}\left( D_{(l+\underline{l})}(l+\underline{l}),(l-\underline{l}%
)\right) \\
&=&-\frac{1}{8}\left( -(D_{(l+\underline{l})}l,\underline{l})+(D_{(l+%
\underline{l})}\underline{l},l)\right) \\
&=&-\frac{1}{8}\left( -(D_{l}l,\underline{l})-(D_{\underline{l}}l,\underline{%
l})+(D_{l}\underline{l},l)+(D_{\underline{l}}\underline{l},l)\right) \\
&=&\Omega -\underline{\Omega }
\end{eqnarray*}

\[
k_{{\bf BB}}^{any}=-(D_{{\bf B}}{\bf B},{\bf B})=0 
\]

\begin{eqnarray*}
where &:&A,B\in \{1,2\},\text{the indices for vectors tangent to }S^{2} \\
\text{note} &:&\text{{}}B\in \{1,2\}\text{ is not bold to distinguish it
from the bi-normal.}
\end{eqnarray*}

In the above, use is made of the fact that $l\cdot l=0$ and $\underline{l}%
\cdot \underline{l}=0$; i.e., they are null vectors. Also, $\underline{l}%
\cdot l=-2$.

\qquad Now, for the case of the $k_{ij}^{\prime }$ $\ $($l^{\prime },%
\underline{l}^{\prime }$), one obtains (using the Ricci rotation
coefficients): 
\begin{eqnarray*}
k_{AB}^{\prime } &=&-\frac{1}{2}(\chi _{AB}^{\prime }+\underline{\chi }%
_{AB}^{\prime }) \\
k_{AN^{\prime }}^{\prime } &=&\epsilon _{A} \\
&& \\
k_{N^{\prime }N^{\prime }}^{\prime } &=&\frac{1}{2}(-\nabla _{N}\ln \phi
+\delta )+\frac{1}{2}(\nabla _{N}\ln \phi +\delta ) \\
&=&\delta
\end{eqnarray*}

These second and last equalities are tautological because $\epsilon _{A}$
and $\delta $ are defined to be $k_{AN}$ and $k_{NN}$ respectively.
Although, these relations do not give us new values for $k_{AN}^{\prime }$
and $k_{NN}^{\prime },$ they do confirm the correctness of the Ricci
coefficients and the previous calculations. Also, note using 
\begin{equation}
tr\underline{\chi }^{\prime }+tr\chi ^{\prime }=2\delta  \label{tr Chi sum}
\end{equation}
(cf. \cite{C&k'93} page 500), one sees that, in an orthonormal basis: $%
trk^{\prime }=\gamma ^{AB}k_{AB}^{\prime }+g_{N^{\prime }N^{\prime
}}\,k_{N^{\prime }N^{\prime }}^{\prime })=(-\delta +\delta )=0$ as expected
for a maximal hypersurface.

In the case of $k_{ij}^{g}$ , (the geodesic null pair) one obtains: 
\begin{eqnarray}
k_{AB}^{g} &=&-\frac{1}{2}(\chi _{AB}+\underline{\chi }_{AB})  \label{KgAB}
\\
&=&-\frac{1}{2}(a^{-1}\,\chi _{AB}^{\prime }+a\,\underline{\chi }%
_{AB}^{\prime })  \label{KgAB2} \\
k_{AN}^{g} &=&W_{A}=\epsilon _{A}+\nabla \hspace{-0.13in}/_{A}\ln a
\label{KgAN} \\
k_{NN}^{g} &=&\Omega +\underline{\Omega }  \label{KgNN} \\
&=&0+\frac{a}{2}(\nabla _{N}\ln \phi +\delta )-\frac{1}{2}D_{\underline{l}%
^{\prime }}a  \nonumber
\end{eqnarray}

\section{Fall-off Laws for Curvature Near Null Infinity}

To complete our task of generalizing and understanding linear and angular
momentum, we will need the expansion for large $r$ (defined in section two)
of the $k_{ij}^{\prime }s$. To do this expansion, we need, in turn, the
decay laws for $a$, $\phi $ and $\delta \,$and $\epsilon _{A}$. Using the
results on page 510 \cite{C&k'93} we write: 
\begin{eqnarray}
\lim_{C_{u,t\,\rightarrow \infty }}r^{2}\,\nabla _{N}\ln \phi &=&\Omega
_{\phi }  \label{phi decay law} \\
\lim_{C_{u,t\,\rightarrow \infty }}r\ln a &=&\Psi ^{\prime }
\label{a decay law}
\end{eqnarray}

$\Omega _{\phi }$ is a function on the $S^{2}$ surface and a function of the
retarded time, $u.$ As a caution against confusion, note that this symbol, $%
\Omega _{\phi }$ , is not the Ricci coefficient $\Omega $. $\Psi ^{\prime }$
is only a function on the $S^{2}$ surface. Both functions are defined in
appendix II, which is taken from page 504 of \cite{C&k'93}.

This implies using equation \ref{a decay law} and appendix III of this
paper: 
\begin{eqnarray*}
D_{\underline{l}^{\prime }}a &=&D_{a\underline{l}}a \\
&=&\left( 1+\frac{\Psi ^{\prime }}{r}\right) D_{\underline{l}}\left( \frac{%
\Psi ^{\prime }}{r}\right) \\
&=&-\frac{\Psi ^{\prime }}{r^{2}}D_{\underline{l}}r+\,\,o(r^{-2}) \\
&=&\frac{\Psi ^{\prime }}{r^{2}}
\end{eqnarray*}

Also, we have the following two decay laws (pg 508 \cite{C&k'93}) taken in
an orthonormal basis:

\begin{equation}
\lim_{C_{u_{,}t\,\rightarrow \,\infty }}r^{2}\delta =\Omega _{\phi }+\Psi
^{\prime }\equiv \delta ^{(2)}  \label{delta2}
\end{equation}

\[
\lim_{C_{u_{,}t\,\rightarrow \,\infty }}r^{2}\epsilon _{A}\equiv E_{A} 
\]

Also, note:

Using equation \ref{tr Chi sum} one gets: 
\begin{eqnarray*}
tr\underline{\chi }^{\prime }+tr\chi ^{\prime } &=&2\delta \\
(\frac{2}{r}+\frac{H^{\prime }}{r^{2}})+(-\frac{2}{r}+\frac{\underline{H}%
^{\prime }}{r^{2}}) &=&\frac{2\delta ^{(2)}}{r^{2}}
\end{eqnarray*}

One gets: 
\[
H^{\prime }+\underline{H}^{\prime }=2\delta ^{(2)} 
\]

Hence, for $k^{\prime }$ we have:

\begin{eqnarray*}
k_{AN^{\,\prime }}^{\prime } &\sim &\frac{E_{A}}{r^{2}} \\
k_{N^{\,\prime }N^{\,\prime }}^{\prime } &\sim &\frac{\delta ^{(2)}}{r^{2}}
\\
trk^{\prime } &\sim &k_{A}^{\,\,\,A}+k_{N}^{\,\,\,\,\,N}=0
\end{eqnarray*}
For $k^{g}$ we have: 
\begin{eqnarray*}
k_{AN}^{g} &\sim &\frac{E_{A}}{r^{2}}+\frac{1}{r}\nabla \hspace{-0.13in}/_{A}%
\frac{\Psi ^{\prime }}{r} \\
k_{NN}^{g} &\sim &\frac{\Omega _{\phi }}{r^{2}} \\
tr\,k^{g} &=&k_{\,\,A}^{g\,\,\,\,A}+k_{\,\,N}^{g\,\,\,\,\,N}=\frac{\Psi
^{\prime }}{r^{2}}
\end{eqnarray*}

With the details of the calculation, the last line above becomes: 
\begin{eqnarray*}
tr\,k^{g} &=&k_{\,\,A}^{g\,\,\,\,A}+k_{\,\,N}^{g\,\,\,\,\,N} \\
&\sim &-\frac{1}{2}\left( a^{-1}\,(\frac{2}{r}+\frac{H^{\prime }}{r^{2}}%
)+a\,(-\frac{2}{r}+\frac{\underline{H}^{\prime }}{r^{2}})\right)
+k_{\,\,NN}^{g}\,\,\,g^{NN} \\
&\sim &-\frac{1}{2}\left( (1-\frac{\Psi ^{\prime }}{r})\,(\frac{2}{r}+\frac{%
H^{\prime }}{r^{2}})+\,(1+\frac{\Psi ^{\prime }}{r})(-\frac{2}{r}+\frac{%
\underline{H}^{\prime }}{r^{2}})\right) +k_{\,\,NN}^{g}\,\,\,g^{NN}\, \\
&\sim &-\frac{1}{2}\left( \frac{2}{r}+\frac{H^{\prime }-2\Psi ^{\prime }}{%
r^{2}}\,+-\frac{2}{r}+\frac{\underline{H}^{\prime }-2\Psi ^{\prime }}{r^{2}}%
\right) +k_{\,\,NN}^{g}\,\,\,g^{NN} \\
&\sim &-\frac{1}{2}\left( \frac{2\delta ^{(2)}-4\,\Psi ^{\prime }}{r^{2}}%
\right) +k_{\,\,NN}^{g}\,\,\,g^{NN} \\
&\sim &\left( \frac{-\delta ^{(2)}+2\,\Psi ^{\prime }}{r^{2}}\right)
+\,k_{\,\,NN}^{g}\,\,\,g^{NN}
\end{eqnarray*}

The second line makes use of equation \ref{KgAB2} and the expansions for $%
tr\chi $ and $tr\underline{\chi }$. The last line is kept in a suggestive
form for use in calculating the linear momentum, which is our next step.

\section{Generalizing the Linear Momentum}

\qquad The question before us now is how to generalize the standard ADM
(spatial) definition of linear momentum (cf. page 11 of \cite{C&k'93}): 
\begin{equation}
P_{i}^{\prime }(ADM)=\frac{1}{8\pi }\lim\Sb \Sigma _{t}  \\ r\rightarrow
\,\infty  \endSb \int (k_{iN^{\prime }}^{\prime }-tr\,k^{\prime
}\,g_{iN^{\prime }})d\mu _{\gamma }\,  \label{spatial (ADM) linear momentum}
\end{equation}

Here $\gamma $ is the metric on the $S^{2}$ surface and $d\mu _{\gamma }$ is
an area element on it.

The definition that takes a limit to null infinity, instead of a spatial
limit, is more general, because the null definition includes the spatial as
a special case. Before just changing the limit in $P_{i}^{\prime }(ADM),$
consider that the ADM definition of angular momentum taken to null infinity
does not yield the proper null definition \cite{Rizzi:NullgoestoSpatial}.
The correct definition is obtained if one uses the geodesic null pair
instead of the standard null pair \cite{Rizzi:NullgoestoSpatial}.
Intuitively, this is no big surprise, since the geodesic null pair is
appropriate to the null region whereas the standard pair is appropriate to
the spatial slice.

Hence, by analogy with null angular momentum, one should use the null
geodesic pair and obtain:

\begin{equation}
\frame{ \ \ $P_{i}^{g}(null)=\frac{1}{8\pi }\lim\Sb u=const  \\ r\rightarrow
\,\infty  \endSb \int (k_{\sigma N}^{g}-tr\,k^{g}$\thinspace $g_{\sigma N})$%
\thinspace $e_{i}^{\sigma }d\mu _{\gamma }\,\,$}
\label{genearlized linear momentum}
\end{equation}

We now must show that this equivalent to the Bondi momentum.

\begin{eqnarray}
&=&\frac{1}{8\pi }\lim_{r\rightarrow \,\infty }\int (k_{\sigma
N}^{g}-tr\,k^{g}\,g_{\sigma N})\,g^{\sigma \lambda }\,\nabla _{\lambda
}\,x_{i}d\mu _{\gamma }  \nonumber \\
&=&\frac{1}{8\pi }\lim_{r\rightarrow \,\infty }\int \left(
g^{NN}\,k_{NN}^{g}\,\nabla _{N}\,x_{i}+g^{AA}\,k_{AN}^{g}\nabla \hspace{%
-0.12in}/_{A}x_{i}-tr\,k^{g}\,g_{NN}\,g^{NN}\,\nabla _{N}\,x_{i}\right) d\mu
_{\gamma }  \label{2nd equation in reduction} \\
&=&\frac{1}{8\pi }\lim_{r\rightarrow \,\infty }\int \left( 
\begin{array}{c}
k_{\,NN}^{g}\,\,\partial _{N}\,x_{i}+k_{AN^{\prime }}^{\prime }\frac{1}{r}%
\nabla \hspace{-0.13in}/_{A}^{\,\,\,0}x_{i}+\frac{1}{r}\nabla \hspace{-0.13in%
}/_{A}^{\,\,\,0}\ln a\cdot \frac{1}{r}\nabla \hspace{-0.13in}%
/_{A}^{\,\,\,0}x_{i} \\ 
-\,g_{NN}\,\left( \frac{-\delta ^{(2)}+2\,\Psi ^{\prime }}{r^{2}}%
+\,k_{\,NN}^{g}\,\,g^{NN}\right) \partial _{N}\,x_{i}
\end{array}
\right) \left( r^{2}-rH\right) d\mu _{\gamma ^{0}}  \nonumber \\
&=&\frac{1}{8\pi }\lim_{r\rightarrow \,\infty }\int \left( \frac{E_{A}}{r^{2}%
}\cdot \nabla \hspace{-0.13in}/_{A}^{\,\,\,0}(\frac{x_{i}}{r})-\frac{\ln a}{r%
}\triangle \hspace{-0.12in}/\,^{0}(\frac{x_{i}}{r})-\,\frac{-\delta
^{(2)}+2\,\Psi ^{\prime }}{r^{2}}\,\partial _{N}\,x_{i}\right) r^{2}d\mu
_{\gamma ^{0}}  \nonumber \\
&=&\frac{1}{8\pi }\lim_{r\rightarrow \,\infty }\int \left( -\frac{\nabla 
\hspace{-0.13in}/_{A}^{\,\,\,0}E_{A}}{r^{2}}(\frac{x_{i}}{r})+2\frac{\Psi
^{\prime }}{r^{2}}(\frac{x_{i}}{r})\,-\,\frac{-\delta ^{(2)}+2\,\Psi
^{\prime }}{r^{2}}\,(\frac{x_{i}}{r})\,\right) r^{2}d\mu _{\gamma ^{0}} 
\nonumber \\
&=&\frac{1}{8\pi }\lim_{r\rightarrow \,\infty }\int \left( -\nabla \hspace{%
-0.13in}/_{A}^{\,\,\,0}E_{A}+2\Psi ^{\prime }\,+\delta ^{(2)}-2\,\Psi
^{\prime }\,\right) (\frac{x_{i}}{r})d\mu _{\gamma ^{0}}  \nonumber \\
&&\frame{$\,\,\,P_{i}^{g}=\frac{1}{8\pi }\lim\Sb u=const  \\ r\rightarrow
\,\infty  \endSb \int \left( \left( \Omega _{\phi }+\Psi ^{\prime }\right)
-\nabla \hspace{-0.13in}/_{A}^{\,\,\,0}E_{A}\right) (\frac{x_{i}}{r})d\mu
_{\gamma ^{0}}$ \ }  \label{genearlized linear reduced}
\end{eqnarray}

\begin{eqnarray}
\text{where} &:&\text{\thinspace \thinspace }e_{i}^{\sigma }\,\text{is an
orthonormal basis: }\,\hat{e}_{r},\underbrace{\hat{e}_{\vartheta },\hat{e}%
_{\varphi }} \\
&&\quad \quad \ \ \ \ \ \ \ \ \ \ \ \ \ \ \ \ \ \ \ \ \ \ \ \ \ \ \ \ \ \ \
\ \ \ \ \ \ \ \hat{e}_{A}  \nonumber \\
&&\text{ }\nabla \hspace{-0.13in}/_{A}^{\,\,\,0}\text{ is the covariant
derivative on the unit sphere} \\
&&\text{superscript }0\text{ means ``on the unit sphere''}
\end{eqnarray}

Where $\sigma \in \{r,\theta ,\phi \}$ and $x_{1}=x$,$\,$\thinspace $x_{2}=y$
, $x_{3}=z$ and $\hat{N}^{\prime }=\hat{r}$. The second equality is
explained in appendix IV. appendix V gives various notes of interest about
the choice of the geodesic null pair.

To further reduce the last expression, we will make use of the following
relation true all along null infinity (pg. 508 \cite{C&k'93}): 
\begin{equation}
\nabla \hspace{-0.12in}/\cdot E=P-\overline{P}-\frac{1}{2}\Sigma \cdot \Xi +%
\frac{1}{2}\overline{\Sigma \cdot \Xi }+\triangle \hspace{-0.12in}%
/^{\,0}\Psi -\Psi ^{\prime }-\Omega _{\phi }^{\prime }
\label{Del E relation}
\end{equation}

This equation is obtained from the equation in the reference by making the
substitution $\Xi \rightarrow -\frac{1}{2}\Xi $ ; this change makes the
notation the same as that used in my previous papers. Also, note I have
changed a variable name from $\Omega $ to $\Omega _{\phi }$.

Let us take the $x$ momentum, $P_{x}^{\prime }$, as an example. Substituting
equation \ref{Del E relation} into equation \ref{genearlized linear reduced}
yields: 
\begin{eqnarray}
&=&\frac{k_{1}}{8\pi }\int \left( \left( \Omega _{\phi }+\Psi ^{\prime
}\right) -\nabla \hspace{-0.12in}/^{\,0}\cdot E\right) \left(
Y_{1-1}-Y_{11}\right) d\mu _{\gamma ^{0}} \\
&=&\frac{k_{1}}{8\pi }\int \left( \Omega _{\phi }+\Psi ^{\prime }-P+%
\overline{P}+\frac{1}{2}\Sigma \cdot \Xi -\frac{1}{2}\overline{\Sigma \cdot
\Xi }-\triangle \hspace{-0.12in}/^{\,0}\Psi +\Psi ^{\prime }+\Omega _{\phi
}^{\prime }\right) \left( Y_{1-1}-Y_{11}\right) d\mu _{\gamma ^{0}} 
\nonumber \\
&=&\frac{k_{1}}{8\pi }\int \left[ \left( \Omega _{\phi }+\Omega _{\phi
}^{\prime }\right) +2\Psi ^{\prime }+\triangle \hspace{-0.12in}/^{\,0}\Psi
^{\prime }-P+\frac{1}{2}\Sigma \cdot \Xi \right] \left(
Y_{1-1}-Y_{11}\right) d\mu _{\gamma ^{0}}  \nonumber \\
&=&I_{1}+I_{2}+I_{3}\text{ }
\end{eqnarray}

Where $k_{1}=\sqrt{\frac{2\pi }{3}}$

In the second from the last equality use is made of $\triangle \hspace{%
-0.12in}/^{\,0}\Psi =-\triangle \hspace{-0.12in}/^{\,0}\Psi ^{\prime }.$ The
first two terms of the integral are defined to be $I_{1}$, the second two
are $I_{2},$ and the third two are $I_{3}$.

Starting with $I_{1}:$%
\[
I_{1}=\frac{1}{8\pi }\int \left( \Omega _{\phi }+\Omega _{\phi }^{\prime
}\right) k_{1}\left( Y_{1-1}-Y_{11}\right) \,d\mu _{\gamma ^{0}} 
\]

Using $\Omega _{\phi }$ and $\Omega _{\phi }^{\prime }$ from page 504 \cite
{C&k'93} 17.0.9 (and making the substitution $u\rightarrow -\,u$ to bring
notation in line as per above) gives: 
\[
\propto \int_{S^{2}}\left( \int_{-\infty }^{\infty }du^{^{\prime
}}\int_{S^{2}}\frac{\overline{\left| \Xi \right| ^{2}\left( u^{\prime
}\right) }}{\mid \hat{x}-\hat{x}^{\prime }\mid }d^{2}x^{\prime }+\frac{1}{2}%
\int_{-\infty }^{\infty }du^{\prime }sgn(-u+u^{\prime })\,\,\overline{\left|
\Xi \right| ^{2}\left( u^{\prime }\right) }\right) \left(
Y_{1-1}-Y_{11}\right) d\mu _{\gamma ^{0}} 
\]

The expansion below implies that each of the terms in the large brackets are
proportional to $Y_{00}.$%
\[
\frac{1}{\mid \hat{x}-\hat{x}^{\prime }\mid }=4\pi \Sigma _{\ell =0}^{\ell
=\infty }\Sigma _{m=-\ell }^{m=\ell }\frac{1}{2\ell +1}Y_{\ell \,,m}^{\ast
}\left( \theta ^{\prime },\varphi ^{\prime }\right) Y_{\ell ,m}\left( \theta
,\varphi \right) 
\]
\[
where:\mid \vec{x}\mid =\mid \hat{x}\mid =1;\text{\thinspace }\mid \vec{x}%
\mid =\mid \hat{x}^{\prime }\mid =1 
\]

Hence: 
\[
I_{1}=0 
\]

\ Now look at$\,I_{2}$: 
\[
I_{2}=k_{1}\int \left( 2\Psi ^{\prime }+\triangle \hspace{-0.12in}%
/^{\,0}\Psi ^{\prime }\,\right) \left( Y_{1-1}-Y_{11}\right) d\mu _{\gamma
^{0}} 
\]

Note that: 
\begin{eqnarray*}
\int \triangle \hspace{-0.12in}/^{\,0}\Psi ^{\prime }\,\left(
Y_{1-1}-Y_{11}\right) d\mu _{\gamma ^{0}} &=&\int \Psi ^{\prime }\triangle 
\hspace{-0.12in}/^{\,0}\left( Y_{1-1}-Y_{11}\right) d\mu _{\gamma ^{0}} \\
&=&\int -2\Psi ^{\prime }\left( Y_{1-1}-Y_{11}\right) d\mu _{\gamma ^{0}}
\end{eqnarray*}
Where use was made of$:\int \triangle \hspace{-0.12in}/\Psi ^{\prime
}f=-\int \nabla \hspace{-0.12in}/_{A}\Psi ^{\prime }\nabla \hspace{-0.12in}%
/^{\,A}f=\int \Psi ^{\prime }\triangle \hspace{-0.12in}/f$

Hence we have: 
\[
\text{I}_{2}=k_{1}\int \left( 2\Psi ^{\prime }-2\Psi ^{\prime }\right)
\left( Y_{1-1}-Y_{11}\right) d\mu =0 
\]

and 
\begin{equation}
P_{x}^{\prime }=\lim_{C_{u_{,}t\,\rightarrow \,\infty }}\frac{k_{1}}{8\pi }%
\int \left( -P+\frac{1}{2}\Sigma \cdot \Xi \right) \left(
Y_{1-1}-Y_{11}\right) d\mu _{\gamma }
\label{linear momentum at null infinity}
\end{equation}

Of course, the remaining components of $P^{\prime }$at null infinity, ($%
P_{y}^{\prime },$ $P_{z}^{\prime })$ follow in the same way.

This equation (\ref{linear momentum at null infinity}) is the same as that
given in \cite{Katz-Lerer96} (also see \cite{Penrose'63},\cite
{Newman-Penrose68}) for the Bondi linear momentum. The Bondi momentum is a
general definition for linear momentum, because it is valid for all of null
infinity and hence also for spatial infinity; that is, it reduces to the ADM
definition in that limit.\footnote{%
This can be seen fairly directly in the case of linear momentum by just
noting that the limits go through essentially unchanged (cf. appendix D of 
\cite{Rizzi:NullgoestoSpatial}) if one takes limits along $\Sigma
_{t}^{\prime }s$ allowing $r\rightarrow \infty .$}

\begin{itemize}
\item  Equation \ref{2nd equation in reduction} above for $P_{i}^{g}$ is
important for its geometrical interpretation; it can be written as: 
\[
P_{i}^{g}=\frac{1}{8\pi }\lim_{r\rightarrow \,\infty }\int \left(
k_{AN}^{g}\nabla \hspace{-0.12in}/_{A}x_{i}+\frac{1}{2}(tr\chi +tr\underline{%
\chi })\,\partial _{N}\,x_{i}\right) \,d\mu _{\gamma } 
\]
\begin{equation}
\frame{$\,\,P_{i}^{g}=\frac{1}{8\pi }\lim_{r\rightarrow \,\infty }\int
\left( W_{A}\nabla \hspace{-0.12in}/_{A}x_{i}+\frac{1}{2}(tr\chi +tr%
\underline{\chi })\,\nabla _{N}\,x_{i}\right) $\thinspace $d\mu _{\gamma
}\,\,$}  \label{Pgeodesic r-part and tanget part}
\end{equation}
\qquad Where $W_{A}$ is the Ricci Coefficient for the null geodesic pair (cf
appendix C of \cite{Rizzi:NullgoestoSpatial}). The geometrical insight that
equation \ref{Pgeodesic r-part and tanget part} contains will be discussed
in section seven.
\end{itemize}

\subsection{Limits to Null and Spatial Infinity}

\bigskip Although the null definition is already a generalized definition,
there is a slightly higher level of generality that can be achieved.
Specifically, it would be nice to be able to say that the new definition of
linear momentum, which uses the geodesic null pair (equation \ref
{genearlized linear momentum}, i.e. $\int (k_{\sigma N}^{g}-tr\,k^{g}$%
\thinspace $g_{\sigma N})$\thinspace $e_{i}^{\sigma }d\mu _{\gamma }\equiv
Q^{g}$) can be taken to null infinity or to spatial infinity and give
respectively the null and ADM spatial definition. This statement is
demonstrated using the fact that the limit to spatial infinity of the
quantity, $\int (k_{\sigma N}^{any}-tr\,k^{any}$\thinspace $g_{\sigma N})$%
\thinspace $e_{i}^{\sigma }d\mu _{\gamma }\equiv Q^{any},$ is invariant
under choice of lapse function $a$ of the null pair chosen in the
generalized curvature.\footnote{%
Consider the invariance under lapse transformation from $k^{\prime }$ to $%
k^{any}$ induced by the lapse function written as $a\,\,\sim \,1+\frac{a1}{r}
$, where $a1=\Psi ^{\prime }$ for the transformation to the geodesic null
pair. The basic invariance arises because the terms involving $a1$ $(\Psi
^{\prime })$ in the first term of $\int (k_{\sigma N}^{any}-tr\,k^{any}$%
\thinspace $g_{\sigma N})$\thinspace $e_{i}^{\sigma }d\mu _{\gamma }$ cancel
with those that appear in the second term. This cancelation for the case of $%
k^{g}$ is seen by following the $a$ terms in equations \ref{2nd equation in
reduction}-\ref{genearlized linear reduced} until the $\Psi ^{\prime }s$
cancel. Invoking the limit direction independence of the decay law (same for
null and spatial limits) shown appendix D of \cite{Rizzi:NullgoestoSpatial},
one sees the lapse transformation invariance for the spatial limit.}
Specifically, we know that the limit of the quantity, $Q^{standard},$ using
the standard null pair gives the standard spatial (ADM) definition of linear
momentum, thus the use of the geodesic pair must yield the same limit
because of the invariance under choice of lapse function.

\section{Generalized Linear and Angular Momentum}

Reference \cite{Rizzi:NullgoestoSpatial} shows that the angular momentum
defined by:

\[
L(\Omega _{(i)})=\frac{1}{8\pi }\lim_{s\rightarrow \infty
}\int_{S^{2}}W_{A}\Omega _{(i)}^{A}dS_{\gamma } 
\]

gives the same result in the spatial and null limit, with the gauge
condition given in \cite{Rizzi:NullgoestoSpatial}, \cite{RizziAMPRL} and 
\cite{RizziMG8}.

Hence, we have the following generalized definitions for angular and linear
momentum:

\begin{eqnarray}
P_{i} &=&\frac{1}{8\pi }\int (k_{\sigma N}^{g}-tr\,k^{g}\,g_{\sigma
N})e_{i}^{\sigma }d\mu _{\gamma }
\label{genearlized linear momentum (w/o limit)} \\
L &=&\frac{1}{8\pi }\int (k_{AN}^{g}-tr\,k^{g}\,g_{AN})\Omega _{(i)}^{A}d\mu
_{\gamma }  \label{genearlized angular momentum (w/o limit)}
\end{eqnarray}

\qquad Where $\Omega _{(i)}^{A}$ is the rotation vector field (see figure 
\ref{Frotation vector fields}) around the $i-axis.$

As stated these reduce to the appropriate limit at spatial and null
infinity. The more universal significance of choosing the null pair
associated with the null geodesics is apparent in these equations. The
significance can be interpreted as arising from the fact that, at null
infinity, it is appropriate to use light rays (which of course travel on
these geodesics) to probe the structure of space-time that has been
established by the mass in the interior of the space-time at some time in
the past. This rich insight can now be explored to understand the physical
significance of the linear and angular momentum defined in equations \ref
{genearlized linear momentum (w/o limit)} and \ref{genearlized angular
momentum (w/o limit)}.

\section{Geometric Picture and Interrelation between L and P}

\subsection{Linear Momentum}

We start our analysis with the linear momentum in the form of equation \ref
{Pgeodesic r-part and tanget part}: 
\[
P_{i}^{g}=\frac{1}{8\pi }\lim_{r\rightarrow \,\infty }\int \left(
W_{A}\nabla \hspace{-0.12in}/_{A}x_{i}+\frac{1}{2}(tr\chi +tr\underline{\chi 
})\,\nabla _{N}\,x_{i}\right) \,d\mu _{\gamma } 
\]

Note that the first term in the integrand is proportional to the torsion
(cf. \cite{RizziThesis}) \ and the second is\ proportional to the net area
increase (cf. \cite{RizziThesis}) . The probing of the space-time structure
by following the path of light rays reveals the meaning of these terms.
Equivalently, each of the terms can be understood by thinking of our
foliation and how it changes along a given direction.

The torsion, $W_{A}=D_{A}(B^{g},N^{g})$, quantifies the degree to which the $%
S^{2}$ foliation surface pops forward (``out of itself'') as one moves
tangentially from a given point (cf. figure \ref{torsion}). That is, the
torsion at point $p\in S^{2}$ quantifies how much the null geodesic
``shifts'' as one moves in the given direction. This shift can, in turn be
interpreted as the linear momentum carried in the given direction. Hence, $%
k_{AN}=W_{A}=\zeta _{A}$ gives the two tangential components of linear
momentum at a given point on the sphere.

To be more complete, one should say that the torsion measures the rate of
twisting (cf. appendix VI) at a given point. For concreteness take the
world-line in Euclidean space-time shown in figure \ref{torsion}; it can be
written as: 
\[
\overrightarrow{x}(t)=\cos \omega t\,\widehat{x}+\sin \omega t\,\widehat{y}+t%
\text{\thinspace }\widehat{t} 
\]

Using the equation for the torsion given in appendix VI, one gets: 
\[
torsion\sim \omega 
\]

Therefore, the torsion is indeed the rate of twisting.

Now, return to our particular case of an $S^{2}$ surface. Moving an
infinitesimal amount in any direction on it is equivalent to moving on a
piece of a circle. Hence, if I move forward in time I am twisting forward;
that is, there is a twist forward that can be interpreted locally on the
foliation as linear momentum which always arises, because of the constraint,
from a certain twist. Looking at the expression of conserved quantities in
terms of angular variable will elucidate the issue. $L\sim I\,\omega \sim
m\,r^{2}\omega $ and $p\sim m\,\omega \,r$. In these terms, the only
difference is the exponent with which the radial coordinate enters the
expression. In any case, it is clear, as stated above, that the torsion is a
measure of the linear momentum in the given tangential direction.

The $\frac{1}{2}(tr\chi +tr\underline{\chi })$ term is the radial term. The
first (respectively, second) term in this expression tells the degree to
which the outgoing (respectively ingoing) null geodesics are expanding.
Hence the sum gives the net momentum carried in the radial direction at a
given point.\footnote{%
Note that, because of the terms that add asymmetrically when one does a
lapse transformation, the tangential and radial terms will not be separately
invariant under such transformations.}

To get the Cartesian component, one must do the correct projections; for the
case considered above (i.e., $P_{x}$), these projections are done with
derivatives of $x$. Finally, we must sum all the contributions to the linear
momentum in the $\widehat{x}$ direction from each small patch of the sphere (%
$S^{2}$surface) to get the total linear momentum of the system at the
retarded time $u.$

An example will help manifest the meaning of the tangential and radial terms
described above. Further, doing the specific calculation, as is often the
case, reveals and elucidates much of the meaning of the terms. We take the
case of a Schwarzschild solution in isotropic coordinates.

\begin{eqnarray}
ds^{2} &=&-\left( \frac{1-\frac{m}{2r}}{1+\frac{m}{2r}}\right)
^{2}dt^{2}+\left( 1+\frac{m}{2r}\right) ^{4}\left( dr^{2}+r^{2}\left\{
d\theta ^{2}+\sin ^{2}\theta d\phi ^{2}\right\} \right)
\label{isotropic coordinates-spherical} \\
&=&-\alpha ^{2}dt^{2}+\alpha _{2}(dx^{2}+dy^{2}+dz^{2})
\label{isotropic coordinates-rectangular} \\
&=&\left( -1+\frac{2M}{r}-\frac{2M^{2}}{r^{2}}\right) dt^{2}  \nonumber \\
&&+\left( 1+\frac{2M}{r}+\frac{3M^{2}}{r^{2}}\right) (dx^{2}+dy^{2}+dz^{2})
\label{isotropic coordinates-expansion}
\end{eqnarray}

where: $\alpha =\frac{1-\frac{m}{2r}}{1+\frac{m}{2r}}=\frac{\alpha _{1}}{%
\alpha _{2}};\,\,\alpha _{2}=1+\frac{m}{2r};\,r=\sqrt{x^{2}+y^{2}+z^{2}}.$
The last line is the metric expanded out to order $O\left( \frac{1}{r}%
\right) ^{2}$, which is the requisite order for linear momentum.

In the Schwarzschild case, one sees that both the tangential and radial
terms are zero. Note, in these coordinates, the constant time, $t$, slice is
maximal $(tr\,k=0)$.

Now, let us boost to a frame at spatial infinity that is moving with
velocity $\overrightarrow{\beta }=\frac{v}{c}\,\widehat{z}$ with respect to
the background field. The intuitive picture is clear; in this new frame the
Schwarzschild black hole is moving at speed $-\beta $ in the $\,\widehat{z}$
direction. However, the choice of coordinate transformation corresponding to
the boost is ambiguous because of the gauge freedom available with the boost.

For simplicity, we will analyze the propagation of the null geodesics of the
boosted Schwarzschild linear momentum on a ${\em spatial}$ slice far from
all mass (i.e., near spatial infinity). On such a slice, the natural null
pair to use is the standard null pair $(l^{\prime },$ $\underline{l}^{\prime
})$. Thus, the tangential and radial terms of the linear momentum are: $%
\epsilon _{A}\nabla \hspace{-0.12in}/_{A}x_{i}$ and $\frac{1}{2}(tr\chi
^{\prime }+tr\underline{\chi }^{\prime })\,\nabla _{N^{\prime }}\,x_{i}$,
respectively.

Calculating even in this simple case is very\ tricky and complicated. In
fact, I have not found this calculation or even one of this type done in any
detail anywhere in the literature. Another paper will remedy this deficiency
in the literature (\cite{RizziCalcLinearMomentum}). Here I will outline the
method. We will consider the lowest order in $\beta .$ First, one {\em cannot%
} use a simple boost in the$\,\widehat{z}$ direction of the form: $t^{\prime
}=\gamma t-\gamma \beta \,z$ and $z^{\prime }=-\gamma \beta \,t+\gamma \,z$.%
\footnote{%
In lowest order in $\beta $, these equations become: $t^{\prime }=t-\beta
\,z $ and $z^{\prime }=-\beta \,t+\,z$ , because we can take $\gamma =1.$}
This simple boosted Schwarzschild gives, in terms of unboosted radius
coordinate: 
\[
g_{\mu \nu }=\left( 
\begin{array}{cccc}
1+\frac{2M}{r}+\frac{3M^{2}}{r^{2}} & 0 & 0 & 0 \\ 
0 & 1+\frac{2M}{r}+\frac{3M^{2}}{r^{2}} & 0 & 0 \\ 
0 & 0 & 1+\frac{2M}{r}+\frac{3M^{2}}{r^{2}} & \frac{4M\beta }{r}+\frac{M^{2}%
}{r^{2}} \\ 
0 & 0 & \frac{4M\beta }{r}+\frac{M^{2}}{r^{2}} & -1+\frac{2M}{r}-\frac{2M^{2}%
}{r^{2}}
\end{array}
\right) 
\]

where $r=\sqrt{x^{\prime 2}+y^{\prime 2}+(z^{\prime }+\gamma \beta t^{\prime
})}$

One must alter the boost transformation with a gauge transformation to make
resulting metric diagonal. The diagonal metric allows one to view the new $%
time$ coordinate, $t\,$, as generating spatial slices ($t=$constant), $%
\Sigma _{t}$, that foliate the space-time with boosted slices. Finding such
a transformation is non-trivial (cf. \cite{RizziCalcLinearMomentum}). The
transformation needed is: 
\begin{eqnarray*}
x &=&xp \\
y &=&yp \\
z &=&zp+\beta \,tp \\
t &=&\left( 1+\frac{4M}{rp}+\frac{9M^{2}}{rp^{2}}\right) \,tp+\beta \text{ }%
zp
\end{eqnarray*}

where $rp=\sqrt{xp^{2}+yp^{2}+zp^{2}}$. Note here that we use $rp$ to label
the gauge boosted ``radius'' coordinate and distinguish it from the simple
boosted (no gauge terms) ``radius'' $r^{\prime }$. To insure the metric is
diagonal, we consider and calculate the momentum on the surface $tp=0$.

In order to calculate the extrinsic curvature $k_{\mu \nu }=$Proj$_{\Sigma
}(D_{\mu }T_{\nu })$, $T$ must be the normal to the leaf of the foliation ($%
t=$constant) of which one wants to know the curvature.\footnote{%
The curvature, of course, is related to the $tr\chi $ and torsion terms
above, and a similar statement applies to the calculation of them. They
require the correct $T^{\prime }$ in the definition of $l^{\prime
}=T^{\prime }-N^{\prime }$. The extrinsic curvature is used as starting
point because its geometric meaning as the curvature as it appears imbedded
in the larger space-time facilitates grasping the importance of the correct
choice of $T.$ This issue is very murky in the literature and so needs to be
emphasized.} Hence, to calculate, for example, the $k$ of a boosted slice in
the boosted frame one must use: $k_{\mu ^{\prime }\nu ^{\prime }}^{\prime }=$%
Proj$_{\Sigma }(D_{\mu ^{\prime }}T_{\nu ^{\prime }}^{\prime })$, where the
prime on the $T^{^{\prime }}$ means use the normal to the boosted slice and
the primes on the indices mean evaluate in components in the boosted frame.%
\footnote{%
It turns out to be easier in many ways to calculate the quantities directly
in the gauge boosted coordinates where we will evaluate the momentum
integral.} The same issue, of course, arises for the related quantities ($%
tr\chi ,\,tr\underline{\chi }$ and the torsion) which we want to calculate.
In particular, as above, one must use the correct $T$ in the definition of $%
l^{\prime }=T^{\prime }-N^{\prime }$. The extrinsic curvature is a helpful
pedagogical starting point \footnote{%
In fact, because one is working on a spatial slice, the first way I did the
calculation of the momentum is by calculating the $k_{ab}$ (where $a,b\in
\{e_{A},N\}$) and then using the expression for $tr\chi +tr\underline{\chi }$
and $\epsilon _{A}$ in terms of $k$ to obtain them.}, because its geometric
meaning as the curvature of a surface as it appears embedded in the larger
space-time facilitates grasping the importance of the correct choice of $T.$
This issue is very murky in the literature and so needs to be emphasized.
Often the literature reads as if one can just do a simple coordinate
transformation to get the $k_{\mu \nu }$ on a second surface from the $%
k_{\mu \nu }$ on a first surface. This is not true even when the surfaces
are related by coordinate transformations as in the above case where one is
moving from a rest surface to a boosted surface.

One now carries out the transformation and does the appropriate
calculations. As a result, for a black hole as viewed from a frame moving at 
$\beta \,\widehat{z}$ (i.e., in the {\em gauge boosted }Schwarzschild
metric) one gets, to first order in $\beta $:

\begin{eqnarray}
\frac{1}{2}(tr\chi ^{\prime }+tr\underline{\chi }^{\prime }) &=&-\frac{%
2\,M\,\beta \cos \theta }{r^{2}}  \label{radial component---boosted Schwarz}
\\
\epsilon _{\theta }\, &=&\frac{2\,M\,\beta \sin \theta }{r^{2}}
\label{tangential component---boosted Schwarz} \\
\nabla _{N^{\prime }}\,z &=&\cos \theta  \nonumber \\
\nabla \hspace{-0.12in}/_{\theta }\,z &\sim &-\sin \theta  \nonumber
\end{eqnarray}

This gives $(\epsilon _{\theta }\nabla \hspace{-0.12in}/_{\theta }\,z$ + $%
\frac{1}{2}(tr\chi ^{\prime }+tr\underline{\chi }^{\prime })\,\nabla
_{N^{\prime }}\,z)\,r^{2}=-2\,M\,\beta \cos ^{2}\theta -2M\beta $ $\sin
^{2}\theta $.\footnote{%
These calculations confirm the statement above, which was based on heuristic
arguments using the motion of the geodesics, that the tangential and radial
components are zero.} Hence, after doing the surface integral and taking the
limit, one gets the following linear momentum as expected:

\[
P_{z}=-M\beta 
\]

\bigskip Note that the radial component (equation \ref{radial
component---boosted Schwarz}) is zero at $\theta =\frac{\pi }{2}$ as one
expects, because the radial motion at those points on the sphere is in the
x-y plane, and hence should not be effected by the $z-$motion. By an exactly
parallel argument, the tangential component (equation \ref{tangential
component---boosted Schwarz}) is zero at $\theta =0,$ because, at this
point, the tangential component is perpendicular to the boost.

\subsection{\protect\bigskip Angular Momentum}

Next the angular momentum is:

\begin{eqnarray*}
L_{(i)} &=&\frac{1}{8\pi }\int (k_{\sigma N}^{g}-tr\,k^{g}\,g_{\sigma
N})\,\Omega _{(i)}^{\sigma }d\mu _{\gamma } \\
&=&\frac{1}{8\pi }\int (k_{AN}^{g}-tr\,k^{g}\,g_{AN})\,\Omega ^{A}d\mu
_{\gamma } \\
&=&\frac{1}{8\pi }\int (W_{A}-\left( \left( -\frac{1}{2}(tr\chi +tr%
\underline{\chi })\,+k_{NN}\right) \,g_{AN}\right) )\,\Omega ^{A}d\mu
_{\gamma }
\end{eqnarray*}

There are several ways to approach this expression; I will choose the most
physical. First, look at the $\widehat{r}$ $(\widehat{N})$ component of the
motion of the null geodesics; in the integrand, it is $\frac{1}{2}(tr\chi +tr%
\underline{\chi })\,$. We decided that this term was the component of linear
momentum at the given location on the $S^{2}$ foliation that is directed
radial outward (i.e. component parallel to $\hat{r}\,$); hence we do not
expect this to contribute to the angular momentum.

This expectation comes from our classical understanding of angular momentum.
Recall classically that $\vec{L}=\vec{r}\times \vec{p}$ or special
relativistically $L^{\mu \nu }=x^{\mu }P^{\nu }-x^{\nu }P^{\mu }$ which
gives the spatial angular momentum is in the z-direction as: $L^{z}\equiv
L^{xy}=xP^{y}-yP^{x}$ and so is the same as the classical in this case and
in this formal sense. In these terms, when $\vec{p}\sim $ $\widehat{r},$ $%
\vec{L}=0.$

As expected the radial component does not contribute because the rotation
vector fields only have components tangent to the given location on $S^{2}.$
From the integrand, $g_{AN}\Omega ^{A}=\Omega \cdot N=0.$ From figure \ref
{Frotation vector fields}, it is obvious that the $\widehat{r}$ is
perpendicular to the rotation vector fields.

The $k_{NN}$ term is a spurious term that arises because of the
irrelevancies introduced by the spatial slice that is not necessary for the
probing done by the light rays at null infinity. What is important is the
spherical foliation induced by the light rays. The rotation vector fields $%
\Omega ^{A}$ take care of this spurious term because they limit us to the
spheres of the foliation.\footnote{%
Recall in the case of linear momentum a cancelation occurred because a $%
k_{NN}$ term appeared in each of the two terms ($k_{iN}$ and $trk$).}

Finally, the term involving $k_{AN}$ $(=W_{A})$ in the integrand above are
the components of the linear momentum in the $e_{A}$ direction, that is the
tangential direction. Hence, we can write the following fashion we can write
the differential contribution of the angular momentum in the integrand as $%
W_{A}\Omega ^{A}$ in analogy to the relation of the classical linear
momentum and angular momentum:

\[
L_{(i)}=\vec{r}\times \vec{p}=p_{A}\,\Omega _{(i)}^{A} 
\]

\section{Conclusion}

We have introduced the generalized linear momentum (equation \ref
{genearlized linear momentum (w/o limit)}) and the generalized angular
momentum (equation \ref{genearlized angular momentum (w/o limit)}). We have
seen that these give (using the foliation described in section two):

\begin{eqnarray*}
P_{i}(Bondi) &=&\lim_{s\rightarrow \infty }\frac{1}{8\pi }\int (k_{\sigma
N}^{g}-tr\,k^{g}\,g_{\sigma N})e_{i}^{\sigma }d\mu _{\gamma } \\
P_{i}(ADM) &=&\lim\Sb r_{laplace}\rightarrow \infty  \\ on\text{ }\Sigma
_{t}  \endSb \frac{1}{8\pi }\int (k_{\sigma N}^{g}-tr\,k^{g}\,g_{\sigma
N})e_{i}^{\sigma }d\mu _{\gamma }
\end{eqnarray*}
\[
and 
\]
\begin{eqnarray*}
L(null) &=&\lim_{s\rightarrow \infty }\frac{1}{8\pi }\int W_{A}\,\Omega
_{(i)}^{A}d\mu _{\gamma } \\
L(ADM) &=&\lim\Sb r_{laplace}\rightarrow \infty  \\ on\text{ }\Sigma _{t} 
\endSb \frac{1}{8\pi }\int W_{A}\,\Omega _{(i)}^{A}d\mu _{\gamma }
\end{eqnarray*}
Using these definitions, we explained that the linear and angular momentum
are found by following the behavior of light rays near null infinity. The
geometry of the ray behavior showed the meaning of the components that
naturally arise in a foliation of the space-time based on the light rays.
The boosted Schwarzschild solution reveals this component behavior in the
way expected. The analysis also elucidated the interrelationship between the
linear and angular momentum allowing one to see the angular momentum can be
understood as an $\vec{r}\times \vec{p}$.

Hence, we have a precise as well as an intuitive understanding momentum in
the far field case where gravity waves are studied. The ideas are relevant
to understanding of signals received by gravitational wave detection
projects such as LIGO, LISA and VIRGO.

I thank Demetrios Christodoulou and James York for reviewing and giving
feedback on this manuscript.

\section{Appendix I: Ricci Rotation Coefficients}

The Ricci Rotation Coefficients are defined below with respect to a null
pair, $e_{3\,,}e_{4}$ (where $e_{3}.$ $e_{4}$ $=-2$ and,$e_{3}=\underline{l}%
,e_{4}=l$), and the spatial vectors, $e_{1},e_{2},$ orthogonal to these two
vectors and tangent to the topologically $S^{2}$ spheres which make up the
foliation. The $e_{\mu }$ are called the null tetrads.

\begin{equation}
\begin{tabular}{cc}
$\chi _{AB}=H_{AB}=\left\langle D_{A}e_{4,}\left. e_{B}\right\rangle \right. 
$ & $\underline{\chi }_{AB}=\underline{H}_{AB}=\left\langle
D_{A}e_{3},\right. \left. e_{B}\right\rangle $ \\ 
$ZZ_{A}=\frac{1}{2}\left\langle D_{3}e_{4},\left. e_{A}\right\rangle \right. 
$ & $\underline{ZZ}_{A}=\frac{1}{2}\left\langle D_{4}e_{3},\left.
e_{A}\right\rangle \right. $ \\ 
$Y_{A}=\frac{1}{2}\left\langle D_{4}e_{4},\left. e_{A}\right\rangle \right. $
& $\underline{Y}_{A}=\frac{1}{2}\left\langle D_{3}e_{3},\left.
e_{A}\right\rangle \right. $ \\ 
$\Omega =\frac{1}{4}\left\langle D_{4}e_{4},\left. e_{3}\right\rangle
\right. $ & $\underline{\Omega }=\frac{1}{4}\left\langle D_{3}e_{3},\left.
e_{4}\right\rangle \right. $ \\ 
$V_{A}=\frac{1}{2}\left\langle D_{A}e_{4},\left. e_{3}\right\rangle \right. $
& 
\end{tabular}
\label{Ricci Definitions}
\end{equation}

Note that the quantity $\Omega $ is not related to the quantity labeled $%
\Omega _{\phi }$ in equation \ref{phi decay law}.

The un-superscripted null pair, $l,\underline{l}$ refer to the (affine)
geodesic pair (so called because $l$ is tangent to a null geodesic);
un-superscripted Ricci rotation coefficients refer to those with respect to
the geodesic null pairs $l,\underline{l}$ . Superscript $t$ on the null pair
refers to $l^{t}=\phi \,l^{\prime },\,\underline{l}^{t}=\phi ^{-1}$ $%
\underline{l}^{\prime }$ (the $S^{2}$ foliation is propagated to be on a
maximal hypersurface $t$). The primed null pair refers to the ``standard''
null pair: $l^{^{\prime }}=T+N=al,\underline{l}^{^{\prime }}=T-N=a^{-1}%
\underline{l}$ where $T$ is the unit normal to the maximal (spatial)
hypersurface, and $N$ is the unit normal to $S_{t,u}$ in the maximal
hypersurface. Ricci rotation coefficients associated with these null tetrads
are distinguished respectively by superscript $t$ and primes.

For the {\em standard} null pair one gets (rearranging \cite{C&k'93} pg.
171):

\begin{eqnarray*}
&&\text{I. \ \ \ \ {\em Standard}}{\em \,\ }\text{null pair }(l^{\prime },%
\underline{l}^{\prime }) \\
&& 
\begin{tabular}{|l|l|}
\hline
$\chi _{AB}^{\prime }=\theta _{AB}-k_{AB}^{\prime }$ & $\underline{\chi }%
_{AB}^{\prime }=-\theta _{AB}-k_{AB}^{\prime }$ \\ \hline
$ZZ_{A}^{\prime }=\nabla \hspace{-0.13in}/_{A}\ln a+\epsilon _{A}$ & $%
\underline{ZZ}_{A}^{\prime }=\nabla \hspace{-0.13in}/_{A}\ln \phi -\epsilon
_{A}$ \\ \hline
$Y_{A}^{\prime }=0$ & $\underline{Y}_{A}^{\prime }=\nabla \hspace{-0.13in}%
/_{A}\ln \frac{\phi }{a}$ \\ \hline
$\Omega ^{\prime }=\frac{1}{2}(-\nabla _{N}\ln \phi +\delta )$ & $\underline{%
\Omega }^{\prime }=\frac{1}{2}(\nabla _{N}\ln \phi +\delta )$ \\ \hline
$V_{A}=\epsilon _{A}=k_{AN}^{\prime }$ &  \\ \hline
\end{tabular}
\end{eqnarray*}

\begin{eqnarray*}
where &:&\text{\thinspace \thinspace } \\
a &=&\frac{1}{\left| \nabla u\right| }=\text{ \{the lapse function of the
foliation } \\
&&\text{induced by u on each }\Sigma _{t}\} \\
\,\,\,\,\theta _{AB} &=&\text{ \{the extrinsic curvature of the surfaces S}%
_{t,u}\text{ } \\
&&\text{relative to }\Sigma _{t}\} \\
k_{ij}^{\prime } &=&\text{\{extrinsic curvature of the maximal slice\}} \\
&=&\frac{1}{2}\text{Spatial Components(}\pounds _{T^{\prime }}\,\tilde{g}%
_{ij})=-(2\phi )^{-1}\partial _{t}\tilde{g}_{ij} \\
\text{Decomposition of }k^{\prime } &:&\text{{}} 
\begin{array}{lll}
\eta _{AB}=k_{AB}^{\prime }\text{,} & \epsilon _{A}=k_{AN}^{\prime }\text{,}
& \delta =k_{NN}^{\prime }
\end{array}
\end{eqnarray*}

One can transforms to a different null pair; this is called a {\em lapse
transformation}:\footnote{%
The physical meaning of this lapse function is discussed in \cite
{RizziThesis}. In general, the lapse function, together with its counter
part, the shift vector may be described as the non-dynamical variable that
tells one how to move forward in time (cf. e.g. \cite{MTW'73}\cite{Wald'84}).%
} and has the form: 
\begin{eqnarray*}
&& 
\begin{array}{ll}
l^{Trans}=a^{-1}l\,, & \underline{l}^{Trans}=a\,\underline{l}
\end{array}
\\
where &:&a\text{ is called the lapse function } \\
&&\text{and is {\em any} function on the }S^{2}\text{ surface.}
\end{eqnarray*}

Note that the normalization $l\cdot \underline{l}=-2$ is preserved under a 
{\normalsize lapse transformation}{\em . }Under the lapse transformation
above, the Ricci Coefficients transform as:

\begin{equation}
\begin{tabular}{cc}
$\chi ^{Trans}=a^{-1}\chi $ & $\underline{\chi }^{Trans}=a\,\chi $ \\ 
$ZZ^{Trans}=ZZ$ & $\underline{ZZ}^{Trans}=\underline{ZZ}$ \\ 
$Y^{Trans}=a^{-2}Y$ & $\underline{Y}^{Trans}=a^{2}\underline{Y}$ \\ 
$\Omega ^{Trans}=\frac{1}{2}a^{-2}\,D_{l}a+a^{-1}\Omega $ & $%
\,\,\,\,\,\,\,\,\,\,\,\,\,\,\,\underline{\Omega }^{Trans}=a\underline{%
\,\Omega }-\frac{1}{2}D_{\underline{l}}a$ \\ 
$V^{Trans}=V+\nabla \hspace{-0.13in}/\ln a$ & 
\end{tabular}
\label{Ricci transformation law}
\end{equation}

Hence, in terms of the {\em geodesic} null pair, $l=a^{-1}l^{\prime },$ $%
\underline{l}=a\underline{l}^{\prime }$ one obtains: 
\begin{eqnarray*}
&&\text{II. \ \ \ \ {\em Geodesic}}{\em \,\ }\text{null pair }(l,\underline{l%
}) \\
&& 
\begin{tabular}{|l|l|}
\hline
$\chi _{AB}=a^{-1}(\theta _{AB}-k_{AB}^{\prime })$ & $\underline{\chi }%
_{AB}=a(-\theta _{AB}-k_{AB}^{\prime })$ \\ \hline
$ZZ_{A}=\nabla \hspace{-0.13in}/_{A}\ln a+\epsilon _{A}$ & $\underline{ZZ}%
_{A}=\nabla \hspace{-0.13in}/_{A}\ln \phi -\epsilon _{A}$ \\ \hline
$Y_{A}=0$ & $\underline{Y}_{A}=a^{2}(\nabla \hspace{-0.13in}/_{A}\ln \frac{%
\phi }{a})$ \\ \hline
$\Omega =-a^{-1}\,\Omega ^{\prime }+a^{-1}\Omega ^{\prime }=0$ & $\underline{%
\Omega }=\frac{a}{2}(\nabla _{N}\ln \phi +\delta )-\frac{1}{2}D_{\underline{l%
}^{\prime }}a$ \\ \hline
$V_{A}\equiv W_{A}=\epsilon _{A}+\nabla \hspace{-0.13in}/\ln a$ &  \\ \hline
\end{tabular}
\end{eqnarray*}

\begin{eqnarray*}
where &:&\text{use is made of }D_{l}a=D_{a^{-1}l^{\prime }}a=-2\,\Omega
^{\prime }\Longrightarrow D_{l^{\prime }}a=-2\,a\Omega ^{\prime } \\
&&\text{(derived using relations from \cite{C&k'93} pg 264)\thinspace
\thinspace }
\end{eqnarray*}

In terms of the $t$-null pair, $l^{t}=\phi \,l^{\prime },$ $\underline{l}%
^{t}=\phi ^{-1}\underline{l}^{^{\prime }}$, one gets: 
\begin{eqnarray*}
&&\text{III. \ \ \ \ }{\em \,t-}\text{{\em null pair\ } }(l^{t},\underline{l}%
^{t}) \\
&& 
\begin{tabular}{|l|l|}
\hline
$\chi _{AB}^{t}=\phi (\theta _{AB}-k_{AB}^{\prime })$ & $\underline{\chi }%
_{AB}^{t}=\phi ^{-1}(-\theta _{AB}-k_{AB}^{\prime })$ \\ \hline
$ZZ_{A}^{t}=\nabla \hspace{-0.13in}/_{A}\ln a+\epsilon _{A}$ & $\underline{ZZ%
}_{A}^{t}=\nabla \hspace{-0.13in}/_{A}\ln \phi -\epsilon _{A}$ \\ \hline
$Y_{A}^{t}=0$ & $\underline{Y}_{A}^{t}=\phi ^{-2}(\nabla \hspace{-0.13in}%
/_{A}\ln \frac{\phi }{a})$ \\ \hline
$\Omega ^{t}=\frac{1}{2}\phi ^{2}\,D_{l^{\prime }}\phi ^{-1}+\,\frac{\phi }{2%
}(-\nabla _{N}\ln \phi +\delta )$ & $\underline{\Omega }^{t}=\frac{\phi ^{-1}%
}{2}(\nabla _{N}\ln \phi +\delta )-\frac{1}{2}D_{\underline{l}^{\prime
}}\phi ^{-1}$ \\ \hline
$V_{A}^{t}=\epsilon _{A}-\nabla \hspace{-0.13in}/\ln \phi =W_{A}-\nabla 
\hspace{-0.13in}/_{A}\ln (a\phi )$ &  \\ \hline
\end{tabular}
\end{eqnarray*}

\section{Appendix II: Definition of Functions, $\Psi $ and $\Omega _{\protect%
\phi }$}

\bigskip $\qquad $The below functions are taken from \cite{C&k'93}, but they
are modified to take into account our current notational definitions. Recall
above we use a definition of $u$ consistent with \cite{ChristoPRL},\cite
{RizziAMPRL},\cite{RizziThesis} and thus different from \cite{C&k'93} as
mentioned in the text of of the current article. The change is made by
taking $\Xi \rightarrow -\frac{1}{2}\Xi $ \ and $u\rightarrow -u$ as needed. 
\begin{eqnarray*}
\Psi &=&-\frac{1}{4\ast 4\pi }\int_{-\infty }^{\infty }du^{^{\prime }}\left(
\int_{S^{2}}\frac{\left| \Xi \right| ^{2}\left( u^{\prime },\hat{x}^{\prime
}\right) }{\mid \hat{x}-\hat{x}^{\prime }\mid }d^{2}x^{\prime }\right) \\
\Psi ^{\prime } &=&\frac{1}{4\ast 4\pi }\int_{-\infty }^{\infty
}du^{^{\prime }}\left( \int_{S^{2}}\frac{\left| \Xi \right| ^{2}\left(
u^{\prime },\hat{x}^{\prime }\right) -\overline{\left| \Xi \right|
^{2}\left( u^{\prime }\right) }}{\mid \hat{x}-\hat{x}^{\prime }\mid }%
d^{2}x^{\prime }\right)
\end{eqnarray*}

\begin{eqnarray*}
\Omega _{\phi } &=&\frac{1}{4\ast 8\pi }\int_{-\infty }^{\infty
}du^{^{\prime }}\int_{S^{2}}\frac{\left| \Xi \right| ^{2}\left( u^{\prime },%
\hat{x}^{\prime }\right) }{\mid \hat{x}-\hat{x}^{\prime }\mid }%
d^{2}x^{\prime }+\frac{1}{4}\ast \frac{1}{2}\int_{-\infty }^{\infty
}du^{\prime }sgn(-u+u^{\prime })\,\,\left| \Xi \right| ^{2}\left( u^{\prime
},\hat{x}\right) \\
\Omega _{\phi }^{\prime } &=&-\frac{1}{4\ast 8\pi }\int_{-\infty }^{\infty
}du^{^{\prime }}\int_{S^{2}}\frac{\left| \Xi \right| ^{2}\left( u^{\prime },%
\hat{x}^{\prime }\right) -\overline{\left| \Xi \right| ^{2}\left( u^{\prime
}\right) }}{\mid \hat{x}-\hat{x}^{\prime }\mid }d^{2}x^{\prime } \\
&&-\frac{1}{4}\ast \frac{1}{2}\int_{-\infty }^{\infty }du^{\prime
}sgn(-u+u^{\prime })\,\left( \left| \Xi \right| ^{2}\left( u^{\prime },\hat{x%
}\right) -\overline{\left| \Xi \right| ^{2}\left( u^{\prime }\right) }%
\right) \,
\end{eqnarray*}

\[
where:\mid \vec{x}\mid =\mid \hat{x}\mid =1;\text{\thinspace }\mid \vec{x}%
\mid =\mid \hat{x}^{\prime }\mid =1 
\]

\section{Appendix III: Derivative of Areal Radius}

Take as the $S^{2}$ surfaces the ones propagated along $l$ $(\underline{l})$
by $l$ $(\underline{l})$ (the geodesic null pair) so that one has:

\[
\int_{S^{2}}d\mu _{\gamma }=4\pi r^{2} 
\]

Taking $D_{\underline{l}}$ of each side yields: 
\begin{eqnarray}
8\pi \,r\,D_{\underline{l}}r &=&\int_{S^{2}}D_{\underline{l}}d\mu _{\gamma }
\nonumber \\
&=&\int_{S^{2}}tr\,\underline{\chi }\,d\mu _{\gamma } \\
&=&4\pi r^{2}\overline{\,tr\,\underline{\chi }}
\end{eqnarray}

Where overbar means averaged over the solid angle.

Hence, we obtain:

\[
D_{\underline{l}}r=\frac{r}{2}\overline{\,tr\,\underline{\chi }} 
\]

Expanding for large $r$ yields:

\[
D_{\underline{l}}r\sim \frac{r}{2}\left( -\frac{2}{r}+\frac{\overline{%
\underline{H}}}{r^{2}}\right) =-1+\frac{\overline{\underline{H}}}{2\,r} 
\]

\section{Appendix IV: Basis Vectors and Forms}

In this article, I make use of the following set of orthonormal basis vector
set for space portion of the space-time: 
\[
e_{r},e_{\theta },e_{\varphi } 
\]
\[
g_{ij}=diag(1,1,1) 
\]
This gives the following vectors basis and dual vector basis sets: 
\begin{eqnarray*}
\hat{e}_{r} &=&\frac{\partial }{\partial r} \\
\hat{e}_{\theta } &=&\frac{1}{r}\frac{\partial }{\partial \theta } \\
\hat{e}_{\phi } &=&\frac{1}{r\sin \theta }\frac{\partial }{\partial \phi }
\end{eqnarray*}
\begin{eqnarray*}
\omega ^{r} &=&dr \\
\omega ^{\theta } &=&r\,d\theta \\
\omega ^{\phi } &=&r\sin \theta \,d\phi
\end{eqnarray*}

In coordinates, one uses the following notation in spherically symmetric
systems: 
\[
ds^{2}=dr^{2}+r^{2}d\theta +r^{2}\sin ^{2}\theta \,d\varphi 
\]

with the basis vectors: 
\[
\partial _{r},\partial _{\theta },\partial _{\phi } 
\]

If one has a contravariant vector, $V,$ in Cartesian coordinates: 
\[
V=V^{\mu \,cart}\,\hat{x}^{\mu } 
\]
In spherical coordinates: 
\[
V=V^{\mu \,\,sphere}\vec{x}_{\mu }^{r} 
\]
In an orthonormal basis: 
\begin{equation}
V=V^{\mu \,\,ortho}\hat{e}_{\mu }  \label{V in orthonormal basis}
\end{equation}
The subscript on $e_{\mu }$ indicates which the basis vector of the
orthonormal set to use. \ To write the $x$-component of $V$ in terms of the
orthonormal components, one can dot equation \ref{V in orthonormal basis}
with $\hat{x}$; this is the same as contracting with $dx$. 
\begin{eqnarray}
V_{x}^{cart} &=&V^{\mu \,\,ortho}<\hat{e}_{\mu },dx>  \nonumber \\
&=&V^{\mu \,\,ortho}dx(\hat{e}_{\mu })  \nonumber \\
&=&V^{\mu \,\,ortho}\,\nabla _{\hat{e}_{\mu }}x  \nonumber \\
&=&V^{\mu }\,\nabla _{\mu }x  \label{V decomposed as deriv of x}
\end{eqnarray}

In the last line, $\nabla _{\mu }\,$is defined to mean $\nabla _{\hat{e}%
_{\mu }}.$

Equation \ref{V decomposed as deriv of x} remains the same if one uses any
orthonormal basis:

\begin{eqnarray*}
\hat{B} &=&\hat{T}^{\prime }\,(\frac{\partial }{\partial t})+const\ast \hat{N%
}^{\prime }\,(\frac{\partial }{\partial r}) \\
\hat{N} &=&\hat{N}^{\prime }\,(\frac{\partial }{\partial r})+const\ast \hat{T%
}^{\prime }\,(\frac{\partial }{\partial t}) \\
\hat{e}_{\theta } &=&\frac{1}{r}\frac{\partial }{\partial \theta } \\
\hat{e}_{\phi } &=&\frac{1}{r\sin \theta }\frac{\partial }{\partial \phi }
\end{eqnarray*}

Also, for reference, the decomposition of Cartesian coordinates into $%
Y_{lm}^{\prime }$'s is given below:

$x:x=\,r\sin \theta \cos \varphi =r\,k_{1}\,\left( Y_{1-1}-Y_{11}\right) $%
\[
\text{where }\ k_{1}=\sqrt{\frac{2\pi }{3}} 
\]

$y:\,\,y=r\sin \theta \sin \varphi =rk_{2}\left( Y_{1-1}+Y_{11}\right) $%
\[
\text{where \ \ }k_{2}=-i\sqrt{\frac{2\pi }{3}} 
\]

$z:\,\,z=r\cos \theta =r\,k_{3}\,Y_{10}$%
\[
\text{where }k_{3}=\sqrt{\frac{4\pi }{3}} 
\]

\section{Appendix V: The Geodesic Null Pair}

Several issues to note on the frame based on the geodesic null pair:

\begin{itemize}
\item  It is handy to use the following decomposition of ${\bf B}^{g}$ and $%
N^{g}$ into ${\bf B}^{\prime }(=T^{\prime })$ and $N^{\prime }(=\frac{%
\partial }{\partial r}):$

\begin{itemize}
\item[*]  $N^{g}\sim N^{\prime }+\frac{\Psi ^{\prime }}{r}T^{\prime }$ ;

\item[*]  ${\bf B}^{g}\equiv T^{g}\sim {\bf B}^{\prime }-\frac{\Psi ^{\prime
}}{r}N^{\prime })$.

\item[*]  Hence, for example, $\partial _{N}\,x_{i}\sim \partial _{N^{\prime
}}\,x_{i}+$ $\frac{\Psi ^{\prime }}{r}\ast \partial _{T^{\prime }}\,x_{i}$ $%
=\partial _{N^{\prime }}\,x_{i}$.
\end{itemize}

\item  Second, using the geodesic null pair:

\begin{itemize}
\item[*]  $e_{A},{\bf B}^{g}$ are surface forming.

\item[*]  $e_{A},l^{g}$ are surface forming.

\item[*]  However, $e_{A},N^{g}$ are surface forming to order O($\left( 
\frac{1}{r}\right) ^{0}$)$.$

\item[*]  Note: $g_{N^{g}N^{g}}=1;g_{{\bf B}^{g}{\bf B}^{g}}=-1.$
\end{itemize}
\end{itemize}

\section{Appendix VI: The geometric meaning of a few Ricci Rotation
Coefficients}

\subsubsection{Shear}

The shear, $\hat{H}_{AB}=\hat{\chi}_{AB}$ ($\underline{\hat{H}}=\underline{%
\hat{\chi}}_{AB}$), corresponds to the traceless part of the extrinsic
curvature of the $S^{2}$ surface, $\chi _{AB}.$ It is the shear of the
outgoing (respectively, ingoing) null rays. Geometrically, the shear tells
one how the shape a small bundle of null rays changes during a short time
period.

\subsubsection{Expansion}

The expansion, $tr(\chi )$, $(tr(\underline{\chi }))$ gives the ``trace''
part of the extrinsic curvature. It tells the amount of expansion of a small
bundle of outgoing (respectively, ingoing) null rays in a short time period.

\subsubsection{Torsion}

For a curve in $R^{3}$, with tangent vector $\hat{v}$, and principle normal $%
\hat{n}=\frac{\partial \hat{v}}{\partial s}/\left| \frac{\partial \hat{v}}{%
\partial s}\right| $,$\,\hat{b}=$ $\hat{v}\times \hat{n}$, the binormal, $\{%
\hat{v},\hat{n},\hat{b}\}$ form an orthonormal set at each point (see figure 
\ref{torsion}). Using the Serret-Frenet Formulae, one defines the torsion as:

\begin{eqnarray*}
\frac{\partial \hat{b}}{\partial s} &=&\text{Torsion}*\,\hat{n}\text{
\thinspace \thinspace \thinspace \thinspace \thinspace or \thinspace
\thinspace \thinspace \thinspace Torsion}=-\frac{\partial \hat{n}}{\partial s%
}\cdot \hat{b} \\
where &:&\text{ s is the natural parameter of the curve(arc length).}
\end{eqnarray*}

Here ``torsion'' measures the degree of twisting out of the initial plane of
motion. In ``2+1'' space-time, this would correspond to the null rays
twisting, like the grooves of a screw, as they move forward in time.

In a similar way, given a spacelike surface in ($g,M)$, with normals $l$ and 
$\underline{l}$, with $\underline{l}$ chosen to be a null geodesic field \ ($%
D\underline{_{l}}\underline{l}=0)$ and $l$ defined such that $l\cdot 
\underline{l}=-2$ and given $b=\frac{1}{2}(l+\underline{l})$, the binormal
defined in reference\cite{Christo-long}, and the principal normal $n=(%
\underline{l}-l)/2$, one defines the torsion, $\zeta :$%
\[
\zeta _{A}=-(D_{e_{A}}n,b)=(D_{e_{A}}b,n)=\frac{1}{2}(D_{e_{A}}l,\underline{l%
}) 
\]

\bibliographystyle{unsrt}
\bibliography{btxdoc,thsisprd}

\pagebreak

\begin{figure}
\caption{The construction of the affine foliation. Let $k,s\rightarrow
\infty $. $C_{\infty }^{-}$ is null infinity. $S_{-\infty }$ is spatial
infinity. Actual $S^{2}$ surfaces $S_{u,\infty }$ foliate null
infinity.}
\label{FAffine Foliation}
\end{figure}
%

\begin{figure}
\caption{A Euclidean worldline shown in one time and two spatial dimensions. Time is
the verticle dimension. The curve is twisting foward in time, so it has non-zero
torsion=$-\frac{\partial \hat{n}}{\partial s}\cdot \hat{b}$.}
\label{torsion}
\end{figure}
%

\begin{figure}
\caption{The three rotation vector fields on a two-sphere}
\label{Frotation vector fields}
\end{figure}
%

\end{document}